\documentclass[aps, prl, reprint, superscriptaddress]{revtex4-1}

\usepackage{amsmath}
\usepackage{amssymb}
\usepackage{wasysym}
\usepackage{graphicx}
\usepackage[hidelinks, hypertexnames=false]{hyperref}
\usepackage{color}
\usepackage{physics}
\usepackage{siunitx}
\usepackage{xcolor}
\usepackage{changes}
\usepackage{comment}
\usepackage{siunitx}
\usepackage{gensymb}
\usepackage{bm}
\usepackage{natbib}
\usepackage{textcomp}

\allowdisplaybreaks

\begin{document}

\title{Ultracold field-linked tetratomic molecules}
\author{Xing-Yan Chen}
\author{Shrestha Biswas}
\author{Sebastian Eppelt}
\author{Andreas Schindewolf}
\affiliation{Max-Planck-Institut f\"{u}r Quantenoptik, 85748 Garching, Germany}
\affiliation{Munich Center for Quantum Science and Technology, 80799 M\"{u}nchen, Germany}
\author{Fulin Deng}
\affiliation{School of Physics and Technology, Wuhan University, Wuhan, Hubei 430072, China}
\affiliation{CAS Key Laboratory of Theoretical Physics, Institute of Theoretical Physics, Chinese Academy of Sciences, Beijing 100190, China}
\author{Tao Shi}\email{e-mail: tshi@itp.ac.cn}
\author{Su Yi}
\affiliation{CAS Key Laboratory of Theoretical Physics, Institute of Theoretical Physics, Chinese Academy of Sciences, Beijing 100190, China}
\affiliation{CAS Center for Excellence in Topological Quantum Computation \& School of Physical Sciences, University of Chinese Academy of Sciences, Beijing 100049, China}
\affiliation{Peng Huanwu Collaborative Center for Research and Education, Beihang University, Beijing 100191, China}
\author{Timon A. Hilker}
\affiliation{Max-Planck-Institut f\"{u}r Quantenoptik, 85748 Garching, Germany}
\affiliation{Munich Center for Quantum Science and Technology, 80799 M\"{u}nchen, Germany}
\author{Immanuel~Bloch}
\affiliation{Max-Planck-Institut f\"{u}r Quantenoptik, 85748 Garching, Germany}
\affiliation{Munich Center for Quantum Science and Technology, 80799 M\"{u}nchen, Germany}
\affiliation{Fakult\"{a}t f\"{u}r Physik, Ludwig-Maximilians-Universit\"{a}t, 80799 M\"{u}nchen, Germany}
\author{Xin-Yu~Luo} \email{e-mail: xinyu.luo@mpq.mpg.de}
\affiliation{Max-Planck-Institut f\"{u}r Quantenoptik, 85748 Garching, Germany}
\affiliation{Munich Center for Quantum Science and Technology, 80799 M\"{u}nchen, Germany}

\date{\today}

\begin{abstract}  
Ultracold polyatomic molecules offer intriguing new opportunities \cite{Doyle2022} in cold chemistry \cite{Balakrishnan2016,Tang2023}, precision measurements \cite{Hutzler2020}, and quantum information processing \cite{Tesch2002,Albert2020}, thanks to their rich internal structure. However, their increased complexity compared to diatomic molecules presents a formidable challenge to employ conventional cooling techniques. Here, we demonstrate a new approach to create ultracold polyatomic molecules by electroassociation \cite{Quemener2023,Deng2023tetramer} in a degenerate Fermi gas of microwave-dressed polar molecules through a field-linked resonance \cite{Avdeenkov2003,Lassabliere2018,Chen2023}. Starting from ground state NaK molecules, we create around $1.1\times10^3$ tetratomic (NaK)$_2$ molecules, with a phase space density of $0.040(3)$ at a temperature of $\SI{134(3)}{nK}$, more than $3{,}000$ times colder than previously realized tetratomic molecules \cite{Prehn2016}. We observe a maximum tetramer lifetime of $\SI{8(2)}{ms}$ in free space without a notable change in the presence of an optical dipole trap, indicating these tetramers are collisionally stable. The measured binding energy and lifetime agree well with parameter-free calculations, which outlines pathways to further increase the lifetime of the tetramers. Moreover, we directly image the dissociated tetramers through microwave-field modulation to probe the anisotropy of their wave function in momentum space. Our result demonstrates a universal tool for assembling ultracold polyatomic molecules from smaller polar molecules, which is a crucial step towards Bose--Einstein condensation (BEC) of polyatomic molecules and towards a new crossover from a dipolar Bardeen–Cooper–Schrieffer (BCS) superfluid \cite{Baranov2002,Gorshkov2011,Deng2023} to a BEC of tetramers. Additionally, the long-lived FL state provides an ideal starting point for deterministic optical transfer to deeply bound tetramer states  \cite{Lepers2013,Christianen2019,Gacesa2021}. 
\end{abstract}

\maketitle

\section{Introduction}
Molecules exhibit a rich set of internal and external degrees of freedom, which can only be fully controlled under ultracold temperatures ($<\SI{1}{mK}$) \cite{Carr2009,Liu2022}. For example, ultracold molecules prepared in well-defined quantum states allow studying quantum dynamics \cite{Koch2019}, chemical reactions with state-to-state control \cite{Liu2022}, and quantum scattering \cite{Park2023,Chen2023,Tang2023} at an unprecedented level. The highly tunable long-range interactions in dipolar molecules also give rise to novel many-body phenomena \cite{Baranov2012} such as exotic dipolar supersolids \cite{Lassabliere2018} and $p$-wave superfluids \cite{Baranov2002,Gorshkov2011,Deng2023}. Furthermore, ultracold polyatomic molecules have emerged as a powerful platform for various applications including tests of beyond-Standard-Model physics \cite{Hutzler2020}, non-equilibrium dynamics \cite{Liu2023}, and quantum information processing \cite{Tesch2002,Wall2015,Albert2020}, thanks to their additional degrees of freedom compared to diatomic molecules. 

Significant progress has recently been made in the field of molecular cooling, enabling quantum degeneracy in ultracold gases of diatomic dipolar molecules \cite{DeMarco2019,Duda2023,Cao2023}. However, for larger molecules, reaching the ultracold regime remains challenging due to their increased complexity and adverse collisional properties. Direct cooling techniques such as buffer gas cooling \cite{Hutzler2012}, supersonic expansion \cite{Segev2017}, beam deceleration \cite{vandeMeerakker2012}, cryofuges \cite{Wu2017}, and optoelectrical Sisyphus cooling \cite{Prehn2016} have only marginally reached ultracold temperatures. Direct laser cooling has been applied to certain classes of polyatomic molecules \cite{Augenbraun2023}, reaching tens of microkelvin \cite{Vilas2022}. However, laser cooling of larger polyatomic molecules faces a rapid increase in the number of vibrational states limiting efficient photon scattering. Recently, magnetoassociation of ultracold molecules via Feshbach resonances has been extended to triatomic NaK$_2$ molecules in the $\SI{100}{nK}$ regime \cite{Yang2022}, where the molecules inherit the low temperature from the atom--diatomic molecule mixture. However, this technique requires resolvable Feshbach resonances between the collisional partners. For larger, polyatomic molecules, the high number of the intermediate collisional states and their fast loss mechanisms at short range results in a nearly universal collisional loss rate \cite{Bause2023}, preventing the occurrence of such Feshbach resonances. 

Here we demonstrate a novel and general approach to form ultracold polyatomic molecules by electroassociation of smaller polar molecules \cite{Quemener2023,Deng2023tetramer}. We create ultracold tetratomic (NaK)$_2$ molecules from pairs of microwave-dressed fermionic NaK molecules by ramping the microwave field across a field-linked (FL) scattering resonance \cite{Avdeenkov2003,Lassabliere2018,Chen2023}. This approach benefits from the universality of FL resonances and can be applied to any molecule with a sufficiently large dipole moment. We measure a lifetime up to $\SI{8(2)}{ms}$ of our FL tetramers near the dissociation threshold and achieve a phase space density of $0.040(3)$. With microwave-field modulation dissociation after time-of-flight, we directly image the tetramers and reveal the expected anisotropic angular distribution.

\section{Field-linked tetramers}
\begin{figure}
	\centering
	\includegraphics[width=\linewidth]{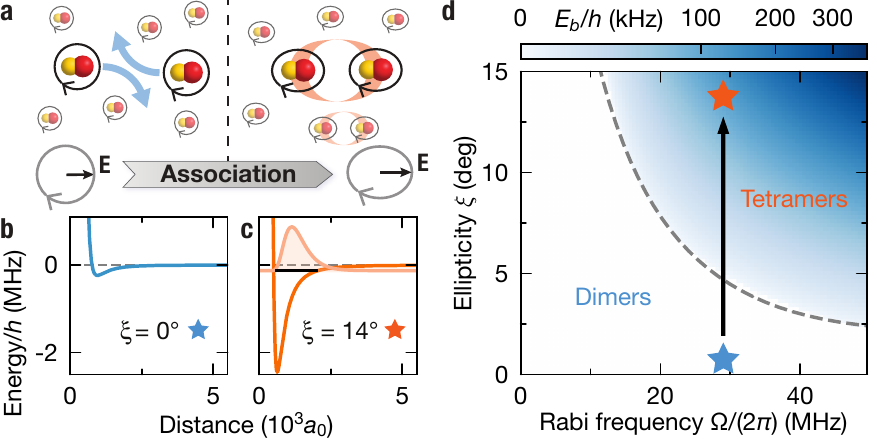}
	\caption{\textbf{Electroassociation of field-linked (FL) tetramers.} \textbf{a}, Microwave dressed NaK dimers are associated into (NaK)$_2$ tetramers as the microwave polarization is ramped from circular to elliptical. \textbf{b,c}, Interaction potentials between two dimers approaching along the long axis of the microwave field at $\xi = 0\degree$ (blue) and $\xi = 14\degree$ (orange). The potential depth increases and a tetramer bound state emerges from the collisional threshold. The light orange line shows the radial wave function of the tetramer, and the black solid line indicates its binding energy. \textbf{d}, Calculated binding energy of the tetramers. The FL resonance (dashed line) marks the onset of the tetramer state. The stars and the arrow mark the electroassociation trajectory in the experiment. Within the range of experimental parameters, there exists only a single FL tetramer state.}
	\label{fig:fig1}
\end{figure}

A microwave FL molecule consists of two microwave dressed polar molecules bound by long-range dipole--dipole interactions. Each constituent molecule is dressed by a near circularly polarized microwave field, which mixes different rotational states and induces a rotating dipole moment of up to $d_0/\sqrt{6}$ in the laboratory frame, where $d_0 \approx \SI{2.7}{Debye}$ is the dipole moment of NaK in its body-fixed frame. The strong induced dipole--dipole interaction potential can host stable tetratomic bound states which give rise to scattering resonances \cite{Chen2023}. By ramping the microwave field across these resonances, a pair of scattering NaK dimers can be adiabatically associated into a (NaK)$_2$ tetramer, as depicted in Fig.~\ref{fig:fig1}a. We refer to this process as electroassociation \cite{Quemener2023}, analogous to magnetoassociation using a magnetic Feshbach resonances \cite{Chin2010}. 

The concept behind electroassociation involves a smooth transition from low-lying scattering states of a dimer pair to the bound tetramer state by gradually ramping the microwave field over time \cite{Quemener2023,Deng2023tetramer}. The increase of the microwave field ellipticity, as depicted in Fig.~\ref{fig:fig1}b,c, enhances the depth of the interaction potential, leading to the emergence of the tetramer state from the collisional threshold and an increase in its binding energy (see Fig.~\ref{fig:fig1}d). Moreover, microwave shielding of the dimers leads to an enhanced collisional stability of the FL tetramers \cite{Karman2018,Anderegg2021,Deng2023tetramer}, which can therefore be efficiently associated from a low entropy gas of dimers.

\section{Binding energy and lifetime}
Our experiments begin with an ultracold gas of optically trapped ($\SI{1064}{nm}$) ground-state $^{23}$Na$^{40}$K molecules with nuclear spin projections $(m_{i,\text{Na}}, m_{i,\text{K}}) = (3/2, -4)$, which are formed from an ultracold atomic mixture by means of magnetoassociation and stimulated Raman adiabatic passage (STIRAP) \cite{Duda2023}. We subsequently dress the molecules with a circularly polarized microwave field, blue detuned to the transition between the ground and the first rotational excited states, in order to shield the molecules from two-body collisions and perform evaporative cooling \cite{Schindewolf2022}. Depending on the trap depth at the end of the evaporation, we prepare various initial conditions of the molecular gas. The minimum temperature is $T = \SI{50(1)}{nK}$ at a dimer molecule number $N_\mathrm{D}$ of $5.7(3)\times10^3$, corresponding to $T/T_\mathrm{F}=0.44(1)$, where $T_\mathrm{F}$ is the Fermi temperature of the trapped gas. The trapping frequencies are $(\omega_{\tilde{x}}, \omega_{\tilde{y}}, \omega_z) = 2\pi\times(42,61,138)\,\mathrm{Hz}$, where $z$ is the vertical direction.

\begin{figure}
	\centering
	\includegraphics[width=\linewidth]{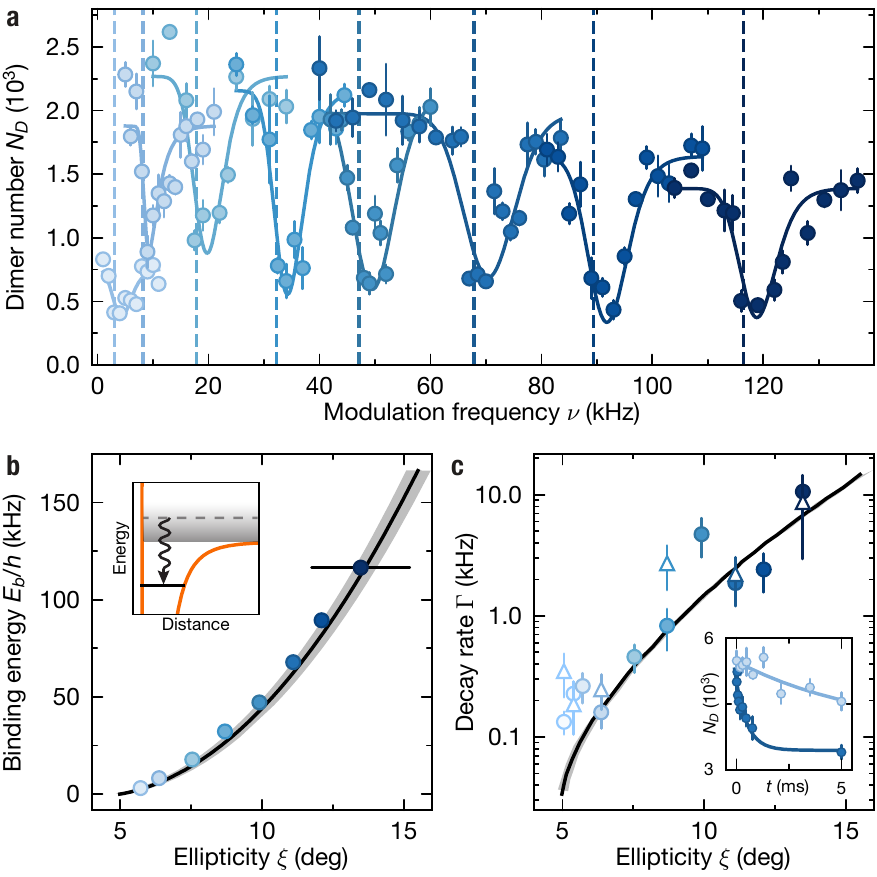}
	\caption{\textbf{Tetramer binding energy and lifetime.} \textbf{a}, Tetramer association spectra at different ellipticities obtained by modulating the ellipticity of the microwave field. The solid lines show the fitted line shape and the dashed lines mark the extracted binding energies. The line shapes are shifted and broadened by the line width of the tetramer states and other technical broadening effects (Methods). The Rabi frequency of the microwave field is $\Omega = 2\pi\times\SI{29(1)}{MHz}$ and detuning $\Delta = 2\pi\times\SI{9.5}{MHz}$. The peak-to-peak modulation amplitude is $1\degree$ and the modulation time is $\SI{100}{ms}$, except for the lowest ellipticity where we use an amplitude of $0.5\degree$ and modulation time of $\SI{400}{ms}$. The error bars represent the standard error of the mean of four repetitions. \textbf{b}, Binding energy $E_\mathrm{b}$ obtained from the association spectra (circles) compared with theory prediction (line). The statistical error bars are smaller than the symbol size. The black error bar marks the systematic uncertainty of ellipticity. The shaded area shows theory calculations including the systematic uncertainty of the Rabi frequency $\Omega$. The inset illustrates the RF association from free to bound states. \textbf{c}, Decay rate $\Gamma$ of the tetramers in time-of-flight (circle) and in trap (triangle), compared to theory calculations (line). The error bars show the fitting errors. The inset shows example decay curves at $\xi = 7(1)\degree$ and $\xi = 11(1)\degree$ in time-of-flight. The error bars represent the standard error of the mean of eight data sets.}
	\label{fig:fig2}
\end{figure}

We probe the binding energy of the tetramers via microwave-field modulation association spectroscopy. We start the experiment with a circularly polarized microwave field at a Rabi frequency $\Omega=2\pi\times\SI{29(1)}{MHz}$ and detuning $\Delta = 2\pi\times\SI{9.5}{MHz}$ \cite{Schindewolf2022}. We then quickly ramp the microwave in $\SI{100}{\micro\second}$ to a target ellipticity $\xi$ above the FL resonance and modulate the ellipticity at various frequencies for up to $\SI{400}{ms}$. The ellipticity $\xi$ is defined such that $\tan\xi$ gives the ratio of the left- and right-handed circularly polarized field components. When the modulation frequency $\nu$ is slightly above the binding energy, tetramers are formed and subsequently decay into lower dressed states accompanied by a large release energy. This leads to a significant reduction of the remaining dimer number, which we detect in the experiment. As shown in Fig.~\ref{fig:fig2}a, we observe clear asymmetric line shapes in the spectra, where the onset frequency of the tetramer association corresponds to the binding energy of the tetramer (Methods). We can thereby determine the binding energy of the tetramers for different target ellipticities (see Fig.~\ref{fig:fig2}b) and find excellent agreement between the experimental data and coupled channel calculations without free parameters (Methods).

Next, we probe the lifetime of the tetramers by measuring their loss dynamics. The dominant loss process for tetramers is spontaneous dissociation into lower microwave dressed states \cite{Karman2018,Deng2023tetramer} accompanied by a large gain in kinetic energy, which, effectively, leads to a one-body decay of the tetramer number. In order to investigate this process, we first create tetramers by ramping the ellipticity to $\xi=8(1)\degree$ in $\SI{0.67}{ms}$, and then quickly ramp to a target ellipticity in $\SI{20}{\micro\second}$. The quick ramp make sure that the measurements at different ellipticities start with the same tetramer and dimer number. There we hold for a variable time, then reverse the ellipticity ramps to dissociate the tetramers back to dimer pairs to map the loss of tetramers during the hold time onto the total dimer number. We turn off the trap after the association to minimize collisional loss. We observe that when the binding energy is high, the observed dimer number quickly undergoes a fast initial decay and afterwards remains constant during the hold time. Near the collisional threshold, the decay is much slower (see inset of Fig.~\ref{fig:fig2}c). These initial decays are much faster than the expected dimer--dimer collisional loss rates, and are absent if we jump from $\xi=0\degree$ to the target ellipticity, so that no tetramers are expected to form. We therefore attribute this initial decay process to the one-body loss of the tetramers, in good agreement with theory predictions (Methods). The corresponding $1/e$ lifetime is longer than $\SI{6(1)}{ms}$ when the binding energy is below $\SI{8.2(2)}{kHz}$, and a maximum of $\SI{8(2)}{ms}$ lifetime is observed near the dissociation threshold. With higher Rabi frequencies and at circular polarization, theory predicts lifetimes in excess of $\SI{100}{ms}$ at $E_\mathrm{b}<h\times\SI{4}{kHz}$ where $h$ denotes the Planck constant (Methods).

To investigate the collisional stability of tetramers, we also assess their lifetimes while the dipole trap remains active. Our observations indicate a combined one-body and two-body loss, and we confirm that the two-body loss arises from dimer--dimer collisions (Methods). Apart from data near the collisional threshold $\xi = 5(1)\degree$, where in-trap measurements are influenced by thermal dissociation, we do not detect notable additional loss of tetramers in in-trap measurements compared to those in time-of-flight experiments. This suggests that tetramers are collisionally stable against collisions with dimers or other tetramers.

\section{Association and dissociation processes}
\begin{figure}
	\centering
	\includegraphics[width=\linewidth]{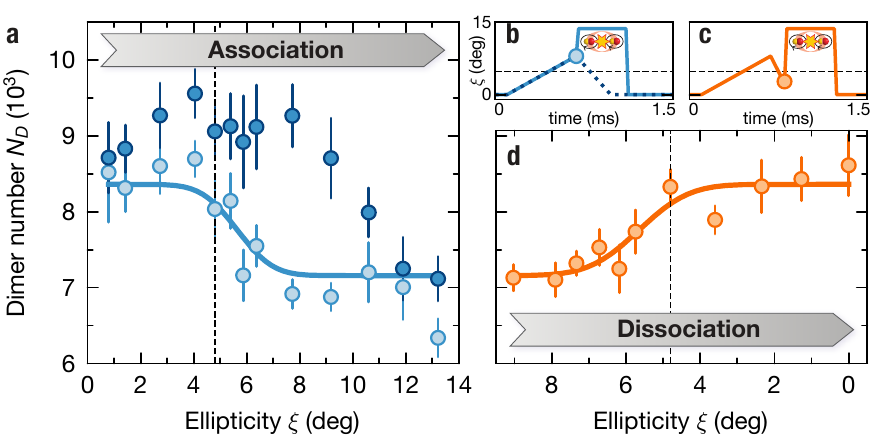}
	\caption{\textbf{Association and dissociation processes.} \textbf{a}, Remaining dimer number $N_\mathrm{D}$ after the association ramp. The dark blue circles show the total number of dimers including dissociated tetramers, while the light circles show the dimer number after removal of the tetramers. The solid blue line is a fit to an error function. The vertical dashed line marks the theoretical resonance position. \textbf{b,c}, Waveform of the association (\textbf{b}) and dissociation (\textbf{c}) ramps. In (\textbf{b}) the blue solid(dashed) line shows the waveform with(without) removal of the tetramers. The horizontal dashed lines indicates the theoretically predicted resonance position. The circles show the target ellipticity of the association or dissociation ramp, which is plotted in (\textbf{a}) and (\textbf{d}), respectively. \textbf{d}, Increase of the detected dimer number during the dissociation ramp. The solid orange line is a fit to an error function. The vertical dashed line shows the predicted resonance position. The error bars represent the standard error of the mean of ten experiment repetitions.}
	\label{fig:fig3}
\end{figure}

We probe the association and dissociation process by ramping the ellipticity starting from $\xi=0$ with a constant ramp speed of $14\degree\,\text{ms}^{-1}$ ($27\degree\,\text{ms}^{-1}$ for the dissociation) to a target ellipticity, as illustrated in Fig.~\ref{fig:fig3}b,c. To distinguish the tetramers from the unpaired dimers we selectively remove the tetramers from the dimer--tetramer mixture by quickly ramping the ellipticity to $\xi=14(1)\degree$ in $\SI{20}{\micro\second}$ and hold for $\SI{0.4}{ms}$. At this point the tetramers are deeply bound and rapidly decay, which removes them from the sample.

Figure~\ref{fig:fig3} shows the evolution of the detected dimer number during the association and the dissociation processes. The number of unpaired dimers (light blue in Fig.~\ref{fig:fig3}a) reduces as we ramp the ellipticity across the FL resonance, indicating tetramer formation. Remarkably, as shown in Fig.~\ref{fig:fig3}d, the number of detected dimers revives as we ramp back to circular polarization, indicating that the formed tetramers can be reversibly dissociated back into dimer pairs. In addition, we characterize the association process without removing the tetramers but followed by a dissociation ramp back to $\xi=0\degree$. The detected dimer number (dark blue in Fig.~\ref{fig:fig3}a) partially revives until $\xi \gtrsim 12\degree$, where the tetramers decay during the ramps before they can be dissociated back into dimers.

\section{Conditions for efficient electroassociation}
\begin{figure}
	\centering
	\includegraphics[width=\linewidth]{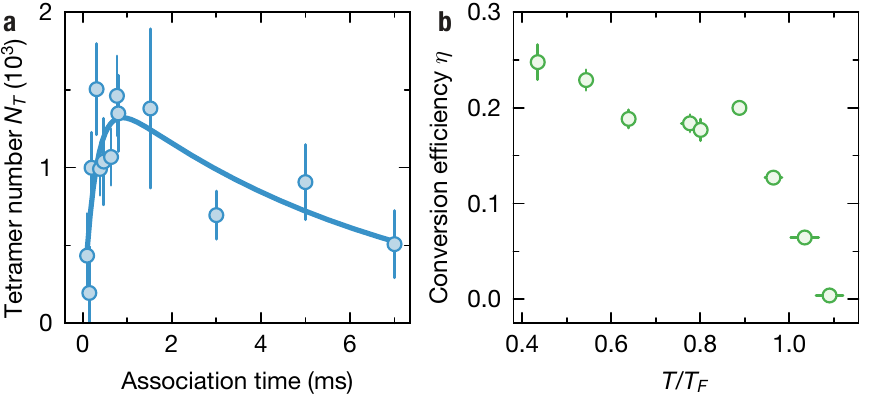}
	\caption{\textbf{Conditions for efficient electroassociation.} \textbf{a}, Tetramer number $N_\mathrm{T}$ as a function of the association time. The solid blue line is a fit to a double exponential function, which captures the formation and decay of the tetramers (Methods). The error bars represent the standard error of the mean of eight repetitions. \textbf{b}, Conversion efficiency $\eta$ as a function of the initial $T/T_\mathrm{F}$ of the dimer gas. We use a ramp speed of $7\degree\,\text{ms}^{-1}$ for the electroassociation. The initial $T/T_\mathrm{F}$ are extracted separately, without performing electroassociation. The error bars represent the standard error of the mean of four repetitions.}
	\label{fig:fig4}
\end{figure}

We move on to identify the optimum condition for electroassociation. We obtain the tetramer number from the difference between images with and without the tetramer removal process outlined previously. First, we probe the timescale of the tetramer formation. We ramp the ellipticity from $\xi = 0(1)\degree$ to $8(1)\degree$ and vary the ramp speed. As shown in Fig.~\ref{fig:fig4}a, we observe the formation of tetramers within $\SI{0.3(1)}{ms}$ and subsequently decay due to the finite lifetime. Assuming the elastic dimer--tetramer scattering rate is on the same order of magnitude as for the dimer--dimer collisions, we estimate that the tetramers scatter in average once during the association, bringing them close to thermal equilibrium with the remaining dimers.

Next, we investigate the role of quantum degeneracy for efficient electroassociation. For magnetoassociation of Feshbach molecules, it has been shown that a low entropy sample is crucial to achieve high conversion efficiency, due the improved phase-space overlap between the atoms \cite{Hodby2005}. Here we vary the degeneracy of our initial dimer samples by changing the final trap depth of the evaporation \cite{Schindewolf2022}. We observe an increase of the conversion efficiency $\eta$, that is the fraction of dimers converted into tetramers, with quantum degeneracy of the dimer gas. We achieve a maximum $\eta = 25(2)\%$ conversion efficiency at $T=0.44(1)T_\mathrm{F}$. Similar as for magnetoassociation \cite{Hodby2005}, a maximum unity conversion efficiency is expected at zero temperature.

\section{Imaging of the dissociated tetramers}

\begin{figure*}
	\centering
	\includegraphics[width=\linewidth]{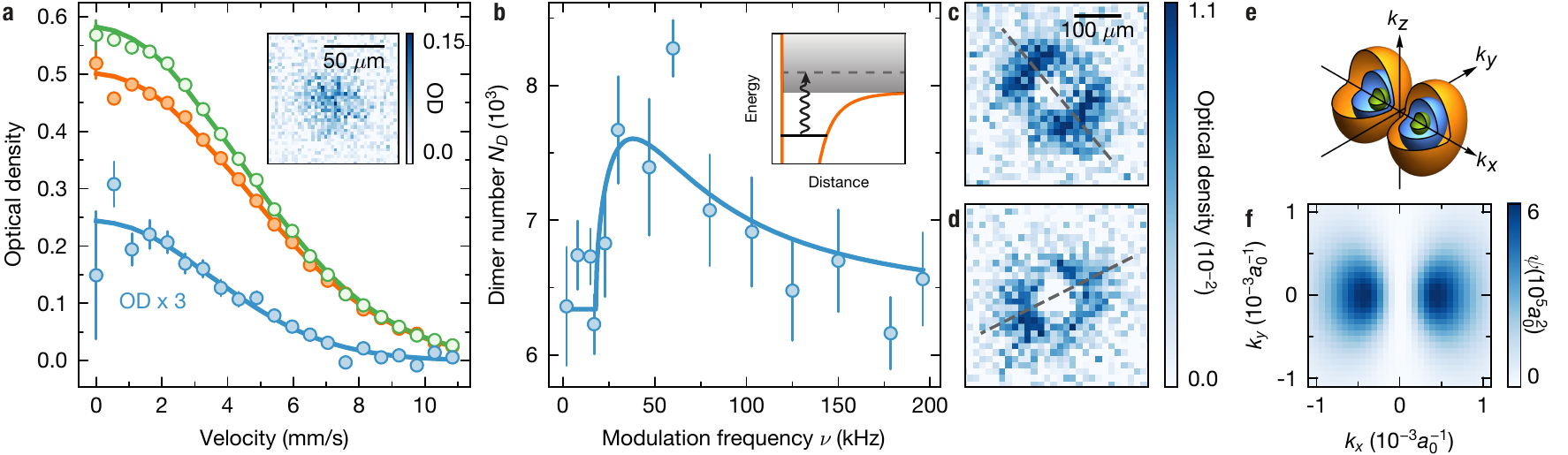}
	\caption{\textbf{Momentum distributions of the dissociated tetramers.} \textbf{a}, Azimuthally averaged optical density (OD) of the samples after ramp dissociation and $\SI{4.5}{ms}$ time-of-flight. The difference (blue) between images with (orange) and without (green) removal of tetramers shows the momentum distribution of the tetramer cloud. The error bars represent standard error of the mean of 60 repetitions. Inset shows the difference image. \textbf{b}, Tetramer dissociation spectrum. We create the tetramers at $\xi=8(1)\degree$ and modulate the ellipticity with an amplitude of $1.4\degree$ for $\SI{2}{ms}$. The solid line is a fit to the dissociation line shape (Methods). The error bars represent the standard error of the mean of ten repetitions. \textbf{c,d}, Time-of-flight images of modulation-dissociated tetramers. We use a modulation frequency of $\SI{30}{kHz}$, with an amplitude of $3.6\degree$ for $\SI{2}{ms}$. While the microwave ellipticity is about the same in (\textbf{c}) and (\textbf{d}) the field orientation differs by about $90\degree$. The dashed lines marks the extracted long axes of the patterns (Methods). The images are averaged over 84 and 40 measurements for (\textbf{c}) and (\textbf{d}), respectively. Each pixel is a binning of $5\times5$ pixels from the raw images. \textbf{e}, Theoretical tetramer wave function in momentum space. The microwave field propagates along the $z$ axis, and its long axis is oriented along the $x$ axis. The cut-open surfaces correspond to probability density of $1.5\times10^8\SI{}{a_0^3}$ (orange), $3.5\times10^8\SI{}{a_0^3}$ (blue), and $6\times10^8\SI{}{a_0^3}$ (green), respectively. \textbf{f}, The theoretical wave function integrated along the propagation axis of the microwave field. The imaging plane  (\textbf{a},\textbf{c}, and \textbf{d}) is roughly perpendicular to the $z$ axis.}
	\label{fig:fig5}
\end{figure*}

We use two methods to obtain absorption images of the tetramers. Firstly, we image the adiabatically dissociated tetramers in time-of-flight to directly probe their temperature. Specifically, we turn off the trap after the electroassociation and image the cloud after $\SI{4.5}{ms}$ of expansion time. To image the molecules, we ramp the ellipticity back to circular to rapidly dissociate the tetramers in $\SI{0.3}{ms}$, then turn off the microwave, reverse the STIRAP to transfer the dimers to the Feshbach-molecule state. Finally we separate the bound atoms via magnetodissociation, directly followed by absorption imaging of the atoms to minimize additional cloud expansion from residual release energy of the tetramer and Feshbach molecule dissociation. The images of the tetramer momentum distribution are obtained by substracting images without from images with removal of tetramers at high ellipticity. Examples of such tetramer images are shown in Fig.~\ref{fig:fig5}a. From a fit to such time-of-flight images and considering the mass of the particles, we determine the temperature of the tetramers to be $\SI{134(3)}{nK}$, which is slightly higher compared to the dimer temperature $\SI{97(6)}{nK}$. The fact that the tetramer cloud is smaller than the dimer background suggests partial thermalization and therefore elastic scattering during the electroassociation. Beyond that, heating might occur during the association and dissociation process. From the number and trapping frequencies, we obtain a peak density of $5.0(2)\times10^{11}\,\text{cm}^{-3}$ and a phase space density of $0.040(3)$ in the trap. We only consider the statistical error in this analysis.

Secondly, we image modulation-dissociated tetramers to probe the angular distribution of their single-particle wave function. A similar protocol has been demonstrated in the photodissociation of diatomic molecules \cite{McDonald2016}. Here we modulate the ellipticity at a modulation frequency $\nu > E_\mathrm{b}/h$, which couples the tetramer states to the scattering continuum. The coupled scattering state possesses a large wave function overlap with the tetramer state, and thus exhibits a similar momentum distribution, which is then probed by time-of-flight imaging. We note that the dissociation pattern is not a one-to-one mapping of the tetramer wave function, but only preserves its angular distribution (Methods). 

We begin by measuring the dissociation spectrum of the tetramers. We create tetramers at $\xi = 8(1)\degree$ via electroassociation, then modulate the ellipticity for $\SI{2}{ms}$ to dissociate tetramers. Meanwhile, we turn off the trap to suppress further association of dimers. Afterwards we remove the remaining tetramers and let the dissociated dimers expand for another $\SI{6}{ms}$ before absorption imaging. The dissociation spectrum, depicted in Figure \ref{fig:fig5}b, demonstrates an increase in the observed dimer number $N_\mathrm{D}$ caused by the presence of dissociated tetramers when the modulation frequency $\nu$ exceeds the frequency associated with the binding energy of the tetramer $E_\mathrm{b}/h = \SI{17.8(3)}{kHz}$. However, at higher frequencies, $N_\mathrm{D}$ declines due to a decrease in dissociation efficiency resulting from the diminished Frank--Condon factor.

We take the difference between images with and without modulation to obtain images of the dissociated tetramers. We verify that modulation at a higher frequency results in a larger pattern due to the higher dissociation energy. We choose a modulation frequency of $\nu = \SI{30}{kHz}$ to optimize the contrast of the images. As shown in Fig.~\ref{fig:fig5}c,d, the dissociation pattern has two lobes, which are oriented along the long axis $x$ of the microwave polarization and match qualitatively with the theoretical wave function in Fig.~\ref{fig:fig5}e,f. Radial integration of the image reveals the angular distribution of the wave function, which follows $p$-wave symmetry \cite{Gaebler2007} in the $p_x$ channel $\cos^2\phi$, where $\phi$ is the angle from the $x$ axis (Methods).  The broken rotational symmetry along the quantization axis is a result of the elliptical microwave polarization. When we rotate the microwave field by roughly $90 \degree$, by flipping the sign of the relative phase between the two feeds of the antenna (Methods), the dissociation pattern is similar but rotated by about $90 \degree$, which demonstrates the tunable control of the tetramer wave function through the microwave field. 

\section{Discussion}
By efficient electroassociation in a degenerate Fermi gas of diatomic molecules, we have created a gas of field-linked tetramers at unprecedentedly cold temperature. The associated (NaK)$_2$ molecules are more than $3{,}000$ times colder than any other tetratomic atomic molecules produced so far~\cite{Prehn2016}. The created tetramers possess a phase space density $11$ orders of magnitude higher the previous record, and is only two orders of magnitude below quantum degeneracy. Remarkably, the lifetime of the long-range FL tetramers is much longer than those observed in polyatomic Feshbach molecules, which are either short lived ($<\SI{1}{\micro\second}$) \cite{Park2023} or unstable in the presence of an optical trap \cite{Yang2022}. These features make them a promising candidate for realizing a BEC of polyatomic molecules.

There are two possible ways to create a BEC of FL tetramers. Firstly, we can make use of the increasing conversion efficiency with lower temperatures. Starting below the critical temperature of $0.14T_\mathrm{F}$, we expect a tetramer BEC to emerge from a degenerate Fermi gas of dimers \cite{Greiner2003}, realizing a BCS--BEC crossover \cite{Zwerger2012} which features anisotropic pairing due to the dipolar interactions \cite{Deng2023}. The other possibility is to extend the tetramer lifetime using the resonance at circular polarization, where the improved shielding increases tetramer lifetime to hundreds of milliseconds. As our experiments suggest that they are collisional stable against dimer–tetramer collisions, it is promising to evaporatively cool tetramers to lower temperatures \cite{Jochim2003}. 

Another interesting direction is to study the excited states of the FL tetramers. As the potential depth increases at higher Rabi frequencies, the interaction potential supports high order FL states, which correspond to excitations of the radial or angular motion of the constituent dimers \cite{Avdeenkov2004}. Such excited FL states have more complex structures, which can be probed similarly with microwave-field modulation.

The creation of FL tetramers opens up a pathway to explore the rich landscape of the four-body potential energy surfaces (PESs). Similar to diatomic molecules, the long-lived weakly bound FL state provides an ideal starting point for deterministic optical transfer to deeply bound states within the PES \cite{Lepers2013,Gacesa2021}. For the PES of (NaK)$_2$ molecules, there are seven energy minima which features distinct geometries including D$_{2h}$, $C_s$, and $C_{2v}$ symmetries \cite{Christianen2019}. These states possess electric dipole and/or quadruple moments, and together with their rich rovibrational structures, opening up new possibilities for studying eight-body collisions and quantum many-body phenomena with both strong dipolar and quadrupolar interactions.

The demonstrated electroassociation via FL resonances is applicable to any polar molecules with a sufficiently large dipole moment \cite{Lassabliere2018,Anderegg2021,Bigagli2023,Lin2023,Quemener2023,Deng2023tetramer}. For example, it can be applied to laser cooled polyatomic molecules to form hexatomic molecules and beyond. Electroassociation can be generalized to d.c. electric fields, where interspecies FL resonances could allow association of two molecules from distinct molecular species. One can even imagine a scalable assembling process, where we sequentially associate pairs of tightly bound molecules into weakly bound FL molecules, convert them into deeply bound states via optical transfer \cite{Lepers2013,Gacesa2021}, then associate these molecules into even larger FL molecules. 

\section{Conclusion}
We have created and characterized field-linked tetratomic (NaK)$_2$ molecules, which is so far the first tetratomic molecules attained in the $\SI{100}{nK}$ regime. The properties of these tetramers are highly tunable with the microwave field, and can be sufficiently long-lived and collisional stable. Thanks to the universality of field-linked resonance, our approach can be generalized to a wide range of polar molecules, including more complex polyatomic molecules. Our results provide a general approach to assemble ultracold polyatomic molecules and open up new possibilities to investigate new quantum many-body phenomena.

\textit{Note}: During completion of this work, we became
aware of a related theoretical proposal on electroassociation of field-linked tetramers from bosonic dimers \cite{Quemener2023}.

\section*{Acknowledgements}
We thank G.\ Quéméner for stimulating discussions. We gratefully acknowledge support from the Max Planck Society, and the Deutsche Forschungsgemeinschaft under Germany's Excellence Strategy -- EXC-2111 -- 390814868 and under Grant No.\ FOR 2247. F.D., T.S., and S.Y. acknowledge support from National Key Research and Development Program of China (Grant No. 2021YFA0718304), and National Natural Science Foundation of China (Grants
No. 11974363 and No. 12274331).

\section*{Author contributions}
All authors contributed substantially to the work presented in this manuscript. X.-Y.C.\ and S.B.\ carried out the experiments and together with S.E.\ and A.S.\ improved the experimental setup. X.-Y.C., S.E., and S.B.\ analyzed the data. F.D., T.S., and S.Y.\ performed the theoretical calculations. T.H., I.B., and X.-Y.L.\ supervised the study. All authors worked on the interpretation of the data and contributed to the final manuscript. 

\bibliography{bibliography}

\clearpage

\renewcommand{\figurename}{Extended Data Fig.}
\setcounter{figure}{0}

\section{Methods}
\subsection{Microwave setup}
The microwave setup is described in detail in Ref.~\cite{Chen2023}. We utilize a dual-feed waveguide antenna capable of synthesizing arbitrary polarization using two independent controllable feeds. The finite ellipticity and interference between these feeds result in an observed change in Rabi frequency of approximately $\pm 4\%$ when adjusting the relative phase, thereby contributing to the systematic uncertainty.

In terms of the control electronics, we have upgraded the amplifiers to $\SI{100}{W}$ in order to achieve higher Rabi frequencies. Additionally, we have implemented filter cavities to suppress phase noise. Furthermore, we have incorporated a voltage-controlled phase shifter, enabling dynamic control of the relative phase between the two feeds for fine-tuning the microwave ellipticity. To maintain a constant output power while adjusting the ellipticity, we monitor the power in each feed using a power detector and employ a feedback control using a voltage-controlled attenuator.

\subsection{Dimer loss near the FL resonance}
\begin{figure}
	\centering
	\includegraphics[width=\linewidth]{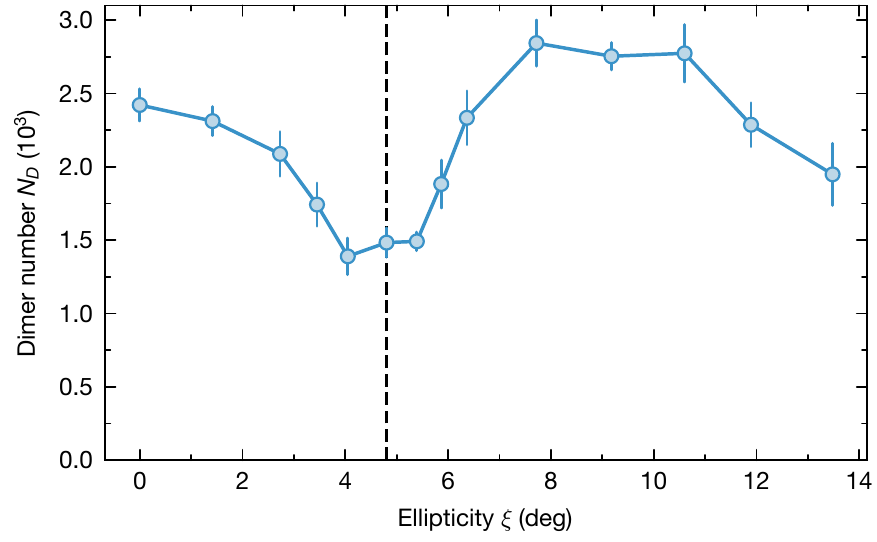}
	\caption{\textbf{Dimer loss near the FL resonance.} Number $N_\mathrm{D}$ of remaining dimers as a function of ellipticity $\xi$. The hold time is $\SI{100}{ms}$. The Rabi frequency of the microwave field is $\Omega =2\pi\times\SI{29(1)}{MHz}$ and the detuning is $\Delta = 2\pi\times\SI{9.5}{MHz}$. The error bars represent the standard error of the mean of four repetitions.}
	\label{fig:ext_resonance}
\end{figure}

We experimentally map out the FL resonance by measuring the dimer loss. Extended Data Figure~\ref{fig:ext_resonance} shows the remained dimer number after a $\SI{100}{ms}$ hold time at $\Omega = 2\pi\times\SI{29(1)}{MHz}$, $\Delta = 2\pi\times\SI{9.5}{MHz}$, as a function of ellipticity $\xi$. The loss dip position matches the theoretical resonance position $\xi = 4.8\degree$.

\subsection{Association spectra analysis}
We assume the dimer loss in the modulation spectra to be proportional to the number of formed tetramers. The line shape can be modeled via Fermi's golden rule \cite{Klempt2008}
\begin{equation}
    N_\mathrm{T}(\nu) \propto \int_{0}^{\infty} \mathrm{d} \epsilon_\mathrm{r} F(\epsilon_\mathrm{r}) g(\epsilon_\mathrm{r}) e^{-(h\nu-E_\mathrm{b}-\epsilon_\mathrm{r})^2/\sigma^2}
\end{equation}
where $\nu$ is the modulation frequency and $E_\mathrm{b}$ is the binding energy of the tetramer. The function $g(\epsilon_\mathrm{r})\propto e^{-\epsilon_\mathrm{r}/k_\mathrm{B} T}$ denotes the number of colliding pairs per relative kinetic energy interval $\mathrm{d}\epsilon_\mathrm{r}$. Here, the temperatures $T$ are obtained from the data located away from the association transitions. The function $F(\epsilon_\mathrm{r}) \propto \sqrt{\epsilon_\mathrm{r}}(1+\epsilon_\mathrm{r}/E_\mathrm{b})^{-2}$ denotes the Franck--Condon factor $F(\epsilon_\mathrm{r})$ between the unbound dimer state and the tetramer state, which we assume to take the same form as for Feshbach molecules \cite{Klempt2008}. The product $F(\epsilon_\mathrm{r})g(\epsilon_\mathrm{r})$ is convoluted with a Gaussian distribution with the width $\sigma$ to account for the line width of the tetramer state and the finite energy resolution. The extracted line width shows a similar trend with ellipticity as the theoretical line width, but slightly larger.

\subsection{Lifetime analysis}
For the measurements in time-of-flight, we verify in absence of tetramers that the two-body loss between dimers is negligible during the hold time. Thus we fit an exponential decay with a constant offset given by the unpaired dimer number $N(t) = 2N_\mathrm{T}e^{-\Gamma t}+N_\mathrm{D}$. The offset $N_\mathrm{D}$ is extracted from the data with ellipticity over $8 \degree$, where the number undergoes a fast initial decay and stay constant afterwards.

For measurements in trap, we ramp up the trap depth by $50\%$ simultaneously with the association, in order to compensate the force from the inhomogenous microwave field. The spatially varying microwave changes the dressed state energy, and thus exerts a force on the molecules which lowers the trap depth and leads to additional loss in the trapped lifetime measurements. 

\begin{figure}
	\centering
	\includegraphics[width=\linewidth]{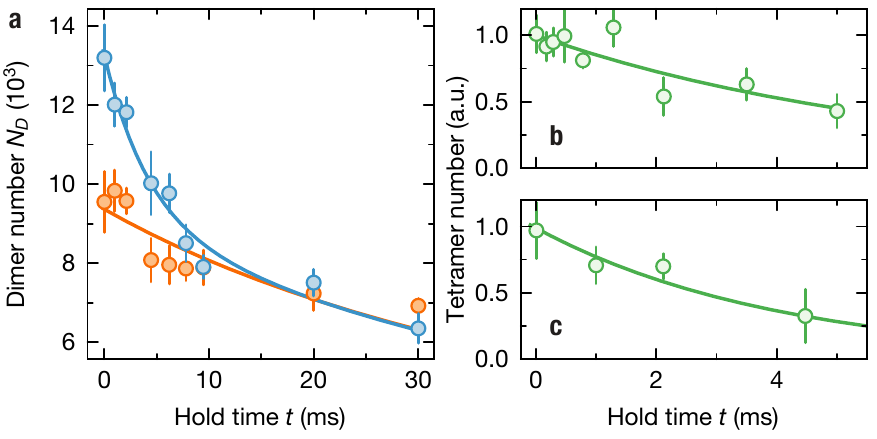}
	\caption{\textbf{Tetramer lifetime in trap and in time-of-flight.} \textbf{a}, Example loss of the molecule number with (orange) and without (blue) removal of tetramers in the trap at $\xi = 7(1)\degree$. \textbf{b}, Normalized tetramer decay measured in time-of-flight at the same ellipticity $\xi$. \textbf{c}, Extracted tetramer number from the data in (\textbf{a}). No notable additional loss is observed compared to (\textbf{b}). The error bars represent standard error of the mean of ten repetitions.}
	\label{fig:ext_lifetime}
\end{figure}

We measure first the total number of tetramer and dimers, and then do a comparison measurement where we remove the tetramers as described in the main text. As shown in Extended Data Fig.~\ref{fig:ext_lifetime}a, we observe a two-body decay in the dimer number, in contrast to the time-of-flight measurements. To account for this background loss, we first determine the two-body loss rate $\Gamma_2$ and initial dimer number $N_\mathrm{D,0}$ from the comparison measurement, then perform a fit of one-body plus two-body decay where we fix $\Gamma_2$ and $N_\mathrm{D,0}$. The fit function is given by $N_D(t) = 2N_\mathrm{T,0}e^{-\Gamma t}+N_\mathrm{D,0}/(1+\Gamma_2 t)$. Extended Data Figure~\ref{fig:ext_lifetime}b,c show the tetramer decay in trap and in free space are similar. 

\subsection{Association timescale analysis}
We apply the following double exponential fit to the tetramer number as a function of ramp time $t$ in Fig.~\ref{fig:fig4}a
\begin{equation}
    N_\mathrm{T}(t) = N_0 (1- e^{-t/\tau})e^{-t_\mathrm{T}/\tau_\mathrm{T}},
\end{equation}
where $\tau$ gives the timescale for association and $\tau_\mathrm{T}$ gives the timescale for tetramer decay. The time $t_\mathrm{T} \approx 0.4(t+t_\mathrm{disso})$ is the time where the ramp is above the FL resonance, which is about a factor of $0.4$ of the association time $t$ and the dissociation time $t_\mathrm{disso} = \SI{0.5}{ms}$. We extract $\tau = \SI{0.3(1)}{ms}$ and $\tau_\mathrm{T} = \SI{2(1)}{ms}$.

\subsection{Hyperfine transitions in the modulation spectra}
\begin{figure}
	\centering
	\includegraphics[width=\linewidth]{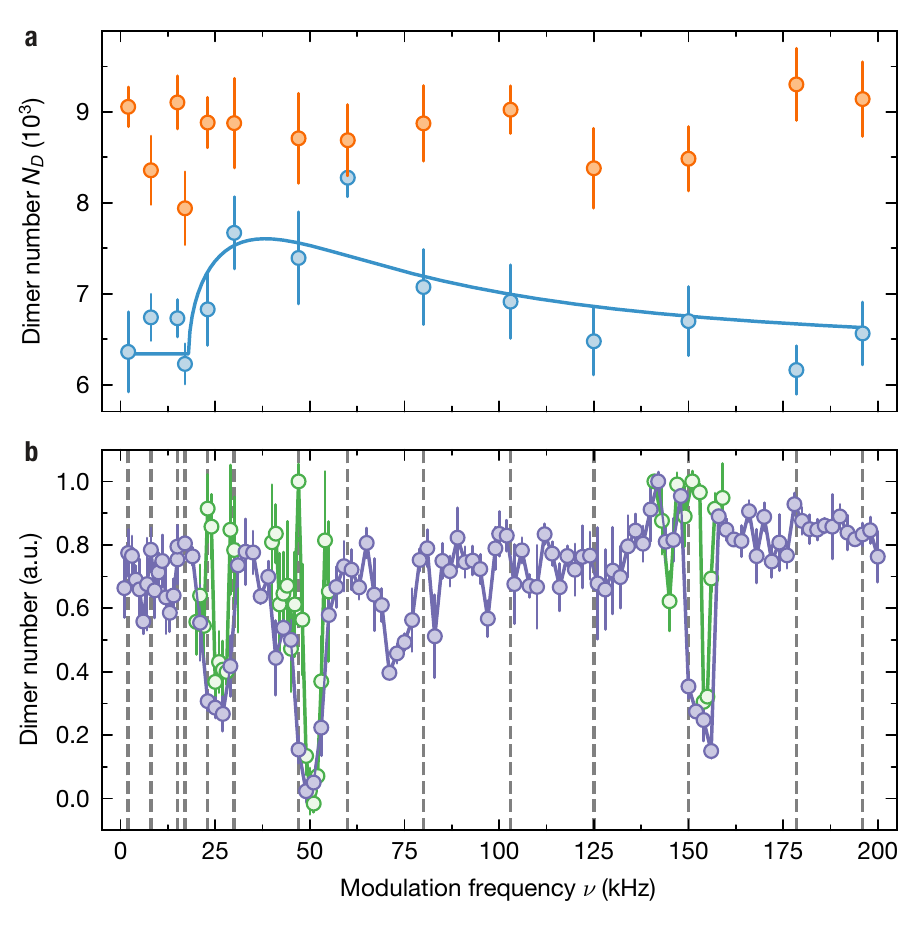}
	\caption{\textbf{Hyperfine transitions of NaK molecules in the modulation spectra.} \textbf{a}, Tetramer dissociation spectrum (blue) compared to background loss (orange). The background loss is measured under the same condition as the dissociation spectrum, except with a fast ramp over the FL resonance so that no tetramers are formed. The absence of loss in the background measurement suggests that the dissociation spectrum is not affected by either the hyperfine transitions or the association. This is ensured by using a small modulation amplitude of $1.4\degree$ over a duration of $\SI{2}{ms}$ and by taking the measurements away from known hyperfine transitions. The error bars represent standard error of the mean of ten repetitions. \textbf{b}, Hyperfine spectrum measured with Landau--Zener sweeps in the modulation frequency and a modulation amplitude of $11\degree$ (purple) and $3.6\degree$ (green). The transitions with the larger modulation amplitude is power broadened compared to the lower amplitude ones. The modulation frequency that we use for the tetramer dissociation spectrum are marked as vertical dashed lines. The error bars represent standard error of the mean of four repetitions.}
	\label{fig:ext_hyperfine}
\end{figure}

We observe effects of hyperfine structure of NaK molecules in the modulation spectra. When we modulate the ellipticity of the microwave by phase modulation, we generate two sidebands that are offset from the carrier by the modulation frequency $\nu$. When $\nu$ matches the ground or excited state hyperfine splitting of the dimer, a two-photon hyperfine transition occurs. In Extended Data Fig.~\ref{fig:ext_hyperfine}b we map out the transition spectrum by Landau--Zener sweeps, where the modulation frequency is ramped from one data point to the next. If a sweep is perfomed over a hyperfine transition, molecules are transferred to another hyperfine state causing a depletion of the detected number of dimers. We observe three major hyperfine transitions from $\SI{2}{kHz}$ to $\SI{200}{kHz}$ and a few weaker ones. We verify that these transitions are not affected by changes in the ellipticity, which confirms that they are not related to the tetramer states. In order to obtain a clear spectrum, when measuring the dissociation spectrum, we employ a small modulation amplitude to minimize power broadening and ensure that we avoid measuring near these transitions.

\begin{figure*}
	\centering
	\includegraphics[width=\linewidth]{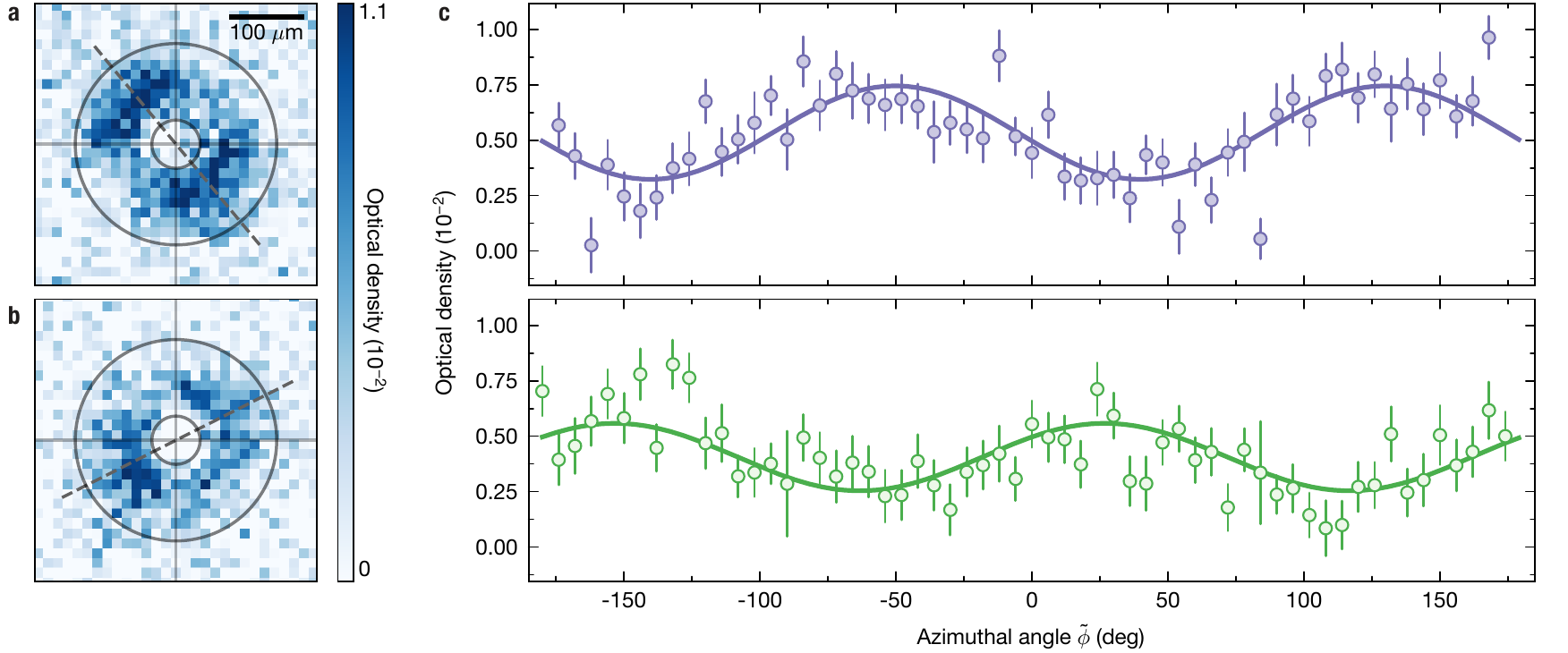}
	\caption{\textbf{Tetramer dissociation patterns and their angular distribution.} \textbf{a,b}, Images of the modulation-dissociated tetramers at different microwave orientations. For each image, we average over the areas between the two circles to obtain the angular distribution of the optical density as shown in \textbf{c}. The dashed line marks the extracted orientation of the microwave, which are $\phi_0 = -50(4)\degree$ (\textbf{a}) and $\phi_0 = 27(4)\degree$ (\textbf{b}). \textbf{c}, Angular distribution of the optical density. The upper(lower) figure corresponds to the image in \textbf{a}(\textbf{b}). 
 }
 \label{fig:ext_rf}
\end{figure*}

\subsection{Tetramer dissociation spectrum analysis}
In addition to hyperfine transitions mentioned above, the association of background dimers into tetramers also affects the measurement of the dissociation spectrum. However, it is worth noting that the association spectra are considerably narrower than the dissociation spectrum, and their influence can be mitigated by employing a small modulation amplitude. To provide evidence for this, we present a comparative measurement in Extended Data Fig.~\ref{fig:ext_hyperfine}a, conducted under identical experimental conditions, except that the ellipticity ramp is as fast as $\SI{0.5}{\micro\second}$ so that no tetramers are formed. Note that the modulation time is much shorter than the association spectra in Fig.~\ref{fig:fig2}a. The observed constant background in this measurement demonstrates that the frequencies at which we measure the dissociation spectrum remain unaffected by hyperfine transitions or association.

We fit the dissociation spectrum with the dissociation line shape similar to Feshbach molecules
\begin{equation}
    N_\mathrm{T}(\nu) \propto \Theta(\nu-E_\mathrm{b}/h)\frac{\sqrt{\nu - E_\mathrm{b}/h}}{\nu^2 + \gamma^2/4} 
\end{equation}
where $\Theta(\nu-E_\mathrm{b}/h)$ is the step function, and $\gamma = \SI{20(7)}{kHz}$ accounts for the broadening of the signal.

\subsection{Angular distribution of the dissociation patterns}
We average along the radial direction of the dissociation patterns to obtain their angular distribution, as shown in Extended Data Fig.~\ref{fig:ext_rf}. The distribution of the average optical density shows a sinusoidal oscillation which matches the $p$-wave symmetry, and we extract the orientation angle $\phi_0$ by a fit to $\propto 1 + c\cos(2(\Tilde{\phi} - \phi_0))$, where $\Tilde{\phi}$ is the angle relative to the horizontal axis of the image and $c$ accounts for the finite contrast.

\subsection{Tetramer lifetime at circular polarization}
The lifetime of tetramers can be improved with better shielding near the circular microwave polarization. With circular polarization, two nearly degenerate tetramer states emerge above the FL resonance at Rabi frequency $\Omega = 2\pi\times\SI{83}{MHz}$ and $\Omega = 2\pi\times\SI{85}{MHz}$, which corresponds to the two $p$-wave channels with angular momentum projection $m = 1$ and $m=-1$, respectively. For the $m=1$ state, the lifetime at binding energy $E_\mathrm{b}<h\times\SI{4}{kHz}$ exceeds $\SI{100}{ms}$. In comparison, we show the decay rate for $\xi=5\degree$ where the resonance occurs at $\Omega = 2\pi\times\SI{28}{MHz}$. For the same binding energy the lifetime is ten times shorter than the $m=1$ state due to the smaller Rabi frequency.

\begin{figure}
	\centering
	\includegraphics[width=\linewidth]{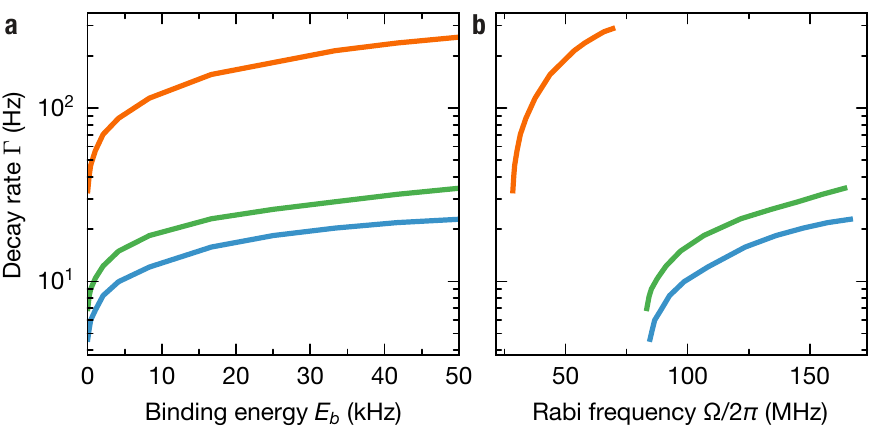}
	\caption{\textbf{Theoretical tetramer decay rate.} Decay rate $\Gamma$ as a function of binding energy (\textbf{a}) and Rabi frequency (\textbf{b}). Each curve in the figures represents a calculation with a fixed $\xi$ while varying the Rabi frequencies, resulting in a range of binding energies. The blue(green) curve corresponds to circular polarization with angular momentum projection $m=1$($m=-1$). The orange curve corresponds to $\xi=5\degree$. 
 }
 \label{fig:ext_circular}
\end{figure}

\subsection{Theory}

We apply the coupled-channel calculation to study the scattering of
molecules governed by the Hamiltonian $\hat{H}=-\nabla ^{2}/M+\sum_{j=1,2}%
\hat{h}_{\mathrm{in}}(j)+V({\mathbf{r}})$, where we take the reduced Planck
constant $\hbar =1$.

The dynamics of a single molecule is described by the Hamiltonian $\hat{h}_{%
\mathrm{in}}=B_{\mathrm{rot}}{\mathbf{J}}^{2}+\frac{\Omega }{2}e^{-i\omega
_{0}t}|\xi _{+}\rangle \langle 0,0|+\mathrm{h.c.}$ with the rotational
constant $B_{\mathrm{rot}}=2\pi \times 2.822\,\mathrm{GHz}$. Here, we only
focus on the lowest rotational manifolds ($J=0$ and $1$) with four states $%
|J,M_{J}\rangle =|0,0\rangle $, $|1,0\rangle $, and $|1,\pm 1\rangle $. The
microwave field of frequency $\omega _{0}$ and elliptic angle $\xi $ couples 
$|0,0\rangle $ and $|\xi _{+}\rangle \equiv \cos \xi \left\vert
1,1\right\rangle +\sin \xi \left\vert 1,-1\right\rangle $ with the
Rabi-frequency $\Omega $. In the interaction picture, the eigenstates of $%
\hat{h}_{\mathrm{in}}$ are $|0\rangle \equiv |1,0\rangle $, $|\xi
_{-}\rangle \equiv \cos \xi \left\vert 1,-1\right\rangle -\sin \xi
\left\vert 1,1\right\rangle $, $|+\rangle \equiv u|0,0\rangle +v|\xi
_{+}\rangle $, and $|-\rangle \equiv u|\xi _{+}\rangle -v|0,0\rangle $, and
the corresponding eigenenergies are $E_{0}=E_{\xi _{-}}=\Delta $ and $E_{\pm
}=(\Delta \pm \Omega _{\mathrm{eff}})/2$, where $u=\sqrt{(1-\Delta /\Omega _{%
\mathrm{eff}})/2}$ and $v=\sqrt{(1+\Delta /\Omega _{\mathrm{eff}})/2}$ with $%
\Delta $ being the detuning and $\Omega _{\mathrm{eff}}=\sqrt{\Delta
^{2}+\Omega ^{2}}$ the effective Rabi frequency.

The interaction of two molecules $V({\mathbf{r}})=V_{\mathrm{dd}}({\mathbf{r}%
})+V_{\mathrm{vw}}({\mathbf{r}})$ contains the dipolar interaction%
\begin{equation}
V_{\mathrm{dd}}({\mathbf{r}})=\frac{d^{2}}{4\pi \epsilon _{0}r^{3}}\left[ 
\hat{\mathbf{d}}_{1}\cdot \hat{\mathbf{d}}_{2}-3(\hat{\mathbf{d}}_{1}\cdot 
\hat{\mathbf{r}})(\hat{\mathbf{d}}_{2}\cdot \hat{\mathbf{r}})\right] ,
\end{equation}%
and the van der Waals interaction $-C_{\mathrm{vw}}/r^{6}$ ($C_{\mathrm{vw}%
}>0$). We can project the Schr\"{o}dinger equation in the two-molecule
symmetric subspace $\mathcal{S}_{7}\equiv \{\left\vert \alpha \right\rangle
\}_{\alpha =1}^{7}=\{|+,+\rangle $, $|+,0\rangle _{s}$, $|+,\xi _{-}\rangle
_{s}$, $|+,-\rangle _{s}$, $|-,0\rangle _{s}$, $|-,\xi _{-}\rangle _{s}$, $%
|-,-\rangle \}$ as $\sum_{\alpha ^{\prime }}\hat{H}_{\alpha \alpha ^{\prime
}}\psi _{\alpha ^{\prime }}({\mathbf{r}})=E\psi _{\alpha }({\mathbf{r}})$,
where $|i,j\rangle _{s}=(|i,j\rangle +|j,i\rangle )/\sqrt{2}$ is the
symmetrization of $|i,j\rangle $. Under the rotating wave approximation, the
Hamiltonian reads%
\begin{equation}
\hat{H}_{\alpha \alpha ^{\prime }}=(-\frac{\nabla ^{2}}{M}+\mathcal{E}%
_{\alpha })\delta _{\alpha \alpha ^{\prime }}+V_{\alpha \alpha ^{\prime }}({%
\mathbf{r}}),
\end{equation}%
where $\mathcal{E}_{\alpha }=\{0,\frac{1}{2}(\delta -\Omega _{\mathrm{eff}}),%
\frac{1}{2}(\delta -\Omega _{\mathrm{eff}}),-\Omega _{\mathrm{eff}},\frac{1}{%
2}(\delta -3\Omega _{\mathrm{eff}}),\frac{1}{2}(\delta -3\Omega _{\mathrm{eff%
}}),-2\Omega _{\mathrm{eff}}\}$ are asymptotic energies of 7 dressed states
with respect to the highest dressed state channel $\left\vert 1\right\rangle 
$, and $V_{\alpha \alpha ^{\prime }}({\mathbf{r}})=\left\langle \alpha
\right\vert V({\mathbf{r}})\left\vert \alpha ^{\prime }\right\rangle $.

To obtain the binding energy and the decay rate of the tetramer in the
dressed state $\left\vert 1\right\rangle $, we consider a pair of
molecules with incident energy $\mathcal{E}_{2}<E<\mathcal{E}_{1}$, the
angular momentum $l$, and its projection $m$ along the $z$-direction. We use
the log-derivative method~\cite{log-Johnson} to numerically solve the Schr\"{%
o}dinger equation in the angular momentum basis, i.e., $\psi _{\alpha }({%
\mathbf{r}})=\sum_{lm}\psi _{\alpha lm}(r)Y_{lm}(\hat{r})/r$, where the loss
induced by the formation of four-body complex is characterized via the
absorption boundary condition at $r=48.5\,a_{0}$. By matching the numerical
solution $\psi _{\alpha lm}(r)$ with the exact wave function in the
asymptotic region $r\rightarrow \infty $, we obtain the scattering
amplitudes $f_{\alpha lm}^{\alpha ^{\prime }l^{\prime }m^{\prime }}$ and the
scattering cross sections $\sigma _{\alpha lm}^{\alpha ^{\prime }l^{\prime
}m^{\prime }}$ from the channel $(\alpha lm)$ to the channel $(\alpha
^{\prime }l^{\prime }m^{\prime })$.

Without loss of generality, we concentrate on the cross section $\sigma
_{210}^{210}$ of the incident and out-going molecules in the channel (210).
When the incident energy is resonant with the tetramer state, a peak appears
in the cross section $\sigma _{210}^{210}$, where the width of the peak is
the decay rate of the tetramer. The cross section $\sigma _{210}^{210}$
quantitatively agrees with the lineshape%
\begin{equation}
\sigma (E)=\frac{2\pi }{k_{2}^{2}}\left\vert ig^{2}G(E)+S_{\mathrm{bg}%
}-1\right\vert ^{2},  \label{sigma}
\end{equation}%
where $G(E)=1/(E-E_{b}+i\Gamma/2)$ is the tetramer propagator, $k_{2}=%
\sqrt{M(E-\mathcal{E}_{2})}$ and $S_{\mathrm{bg}}$ are the incident momentum
and the background scattering amplitude of molecules in the dressed state
channel $|2\rangle $, respectively. By fitting $\sigma _{210}^{210}$ and $%
\sigma (E)$, we obtain the binding energy $E_{b}$ and the decay rate $\Gamma$ of the tetramer. We remark that for the incident and out-going
molecules in other channels {$\alpha =2\sim 7$}, the propagator $G(E)$ in
Eq.~(\ref{sigma}) does not change. Therefore, fitting $\sigma _{\alpha
lm}^{\alpha ^{\prime }l^{\prime }m^{\prime }}$ in a different scattering
channel leads to the same binding energy $E_{b}$ and the decay rate $\Gamma$.

For the tetramer with a small decay rate, its wave function $\psi _{b}(%
\mathbf{r})$ can be obtained via solving the Schr\"{o}dinger equation $H_{%
\mathrm{eff}}\psi _{b}(\mathbf{r})=\bar{E}_{b}\psi _{b}(\mathbf{r})$. The
single-channel model $H_{\mathrm{eff}}=-\nabla ^{2}/M+V_{\mathrm{eff}}({%
\mathbf{r}})$ is determined by the effective potential~\cite{Deng2023}%
\begin{align}
V_{\mathrm{eff}}(\mathbf{r})& =\frac{C_{6}}{r^{6}}\sin ^{2}\theta \left\{ 1-%
\mathcal{F}_{\xi }^{2}(\phi )+[1-\mathcal{F}_{\xi }(\phi )]^{2}\cos
^{2}\theta \right\}   \notag \\
& +\frac{C_{3}}{r^{3}}\left[ 3\cos ^{2}\theta -1+3\mathcal{F}_{\xi }(\phi
)\sin ^{2}\theta \right]   \label{Vps}
\end{align}%
for two molecules in the dressed state channel $\left\vert 1\right\rangle $,
where $\mathcal{F}_{\xi }(\phi )=-\sin 2\xi \cos 2\phi $, $\theta $ and $\phi 
$ are the polar and azimuthal angles of ${\mathbf{r}}$. The strength $%
C_{3}=d^{2}/\left[ 48\pi \epsilon _{0}(1+\delta _{r}^{2})\right] $ of the
negated DDI only depends on the relative detuning $\delta _{r}=|\Delta
|/\Omega $, while the $C_{6}$-term describes an anisotropic shielding
potential that prevents destructive short-range collisions. Using the
B-spline algorithm, we obtain the binding energy $\bar{E}_{b}$ and the
wave function $\psi _{b}(\mathbf{r})\sim Y_{1-}(r)\varphi _{1}(r)/r$ of the
first tetramer bound state, where $Y_{1-}(r)=(Y_{11}(r)-Y_{1-1}(r))/\sqrt{2}$%
. The binding energies $\bar{E}_{b}$ and $E_{b}$ obtained from the
single-channel model and the 7-channel scattering calculation agree with
each other quantitatively for the small $\xi $ and $\Omega $. For the
largest $\xi $ and $\Omega $ in Fig.~\ref{fig:fig1}, the relative error of $%
\bar{E}_{b}$ is less than $30\%$. The tetramer wave function in the momentum
space is the Fourier transform $\psi _{b}(\mathbf{k})=\int d\mathbf{r}e^{-i%
\mathbf{k\cdot r}}\psi _{b}(\mathbf{r})/(2\pi )^{3/2}$ of $\psi _{b}(\mathbf{%
r})$.

For the modulation dissociation, the transition probability $p_{\mathbf{k}}$
to the momentum state $\mathbf{k}$ is determined by the coupling strength $g_{\mathbf{k}}=\int d{%
\mathbf{r}}\psi_{\mathbf{k}}^{\ast }({\mathbf{r}})\partial_{\xi }V_{%
\mathrm{eff}}({\mathbf{r}})\psi_{b}({\mathbf{r}})$. Here, $\psi_{\mathbf{k}}({\mathbf{r}})$ represents the wave function of the scattering state. The coupling strength $g_{\mathbf{k}}$ is primarily influenced by $Y_{1-}(\hat{k})$, which characterizes the angular distribution of $\psi_{b}(\mathbf{k})$. This dominance arises due to the fact that $\partial_{\xi}V_{\mathrm{eff}}$ maintains mirror symmetry with respect to the $x$-$y$ plane. Therefore, by measuring $p_{\mathbf{k}}\sim \left\vert Y_{1-}(\hat{k})\right\vert ^{2}$, we can effectively probe the angular dependence of the tetramer state in momentum space.

\end{document}